\documentclass{article}
\usepackage{graphics}
\usepackage{epsfig}

\begin{document}

\begin{center} {\large \bf Single-atom quantum memory with degenerate atomic levels}
\end{center}
\bigskip
\begin{center}
\textbf{V. A. Reshetov, E. N. Popov}\\
\bigskip
\textit{Department of General and Theoretical Physics, Tolyatti
State University, 14 Belorousskaya Street, 446667 Tolyatti, Russia}
\end{center}

\begin{abstract}
The storage and retrieval of a single-photon polarization q-bit by
means of STIRAP through the atoms with degenerate levels is studied
theoretically for arbitrary polarization of the driving laser field
and arbitrary values of the angular momenta of resonant atomic
levels. The dependence of the probability of long-term photon
storage on the polarization of the driving field and on the initial
atomic state is examined.
\end{abstract}

\section{Introduction}

The search for an efficient optical quantum memory and its
implementation for quantum networks is an active area (see, e.g.,
the reviews \cite{n1,n2,n3,n4}). The role of carrier of information
-- flying q-bit -- in such devices is played by a single photon and
the most natural way to encode the q-bit is provided by its two
polarization degrees of freedom (see \cite{n5,n6,n7,n8} for some
recent proposals). While in the majority of the proposed schemes the
photon quantum state is mapped to the quantum state of some atomic
ensemble \cite{n6,n7,n8}, in the experiment \cite{n5} the photon
polarization q-bit was recorded in the superposition of magnetic
substates of the long-lived degenerate level of a single rubidium
atom. Such single-atom quantum memory offers an advantage of
processing of the stored quantum information. Since at the storage
stage the state of a single atom, in which the photon q-bit is
encoded, is well determined, it may be altered in a controlled way,
e.g., by some laser pulse, so that the state of the retrieved photon
will differ from that of the recorded one in a desirable way
implementing thus the performance of quantum gates. The setup
employed in \cite{n5} for the photon storage and retrieval was based
on the same technique of vacuum stimulated Raman scattering
involving adiabatic passage (STIRAP), which was previously
successfully employed for deterministic single photon emission
\cite{n9,n10,n11}. In these experiments the three-level $\Lambda$ -
type atom was trapped inside the high-finesse cavity, one branch of
the atomic $\Lambda$ - type transitions was coupled to the quantized
cavity field, while the other one was coupled to the driving
coherent laser field. The STIRAP with the three-level atom with
non-degenerate levels is well described in the reviews and textbooks
(see, e.g., \cite{n12,n13}). In case of non-degenerate levels there
exists the only dark state -- the superposition of the ground state
$a$ and some metastable state $b$ -- uncoupled to the excited level
$c$ (Figure 1). In course of Raman scattering the atom is
adiabatically transferred from the initial state $b$ to the target
state $a$. The STIRAP with degenerate atomic levels and with
classical coherent resonant fields was studied in \cite{n14}, were
it was shown that unlike the non-degenerate case the population from
the initial state $b$ is not always totally transferred to the
target level $a$. The objective of the present paper is to analyze
the process of storage and retrieval of a single-photon polarization
q-bit by means of STIRAP through the atoms with degenerate levels
for arbitrary values of the angular momenta of resonant atomic
levels and to study the dependence of the probability of long-term
photon storage on the polarization of the driving field and on the
initial atomic state.

In section 2 the interaction model is described and the
instantaneous eigenvectors of the interaction operator, which
determine the evolution operator in the adiabatic approximation, are
constructed. In case of degenerate levels there appear the new types
of these eigenvectors, non-existing in case of non-degenerate
levels, like the dark states, which atomic part belongs to only one
of the lower levels $a$ or $b$, and the bright states, which couple
the excited level $c$ with only one of the lower levels $a$ or $b$.
In section 3 the formula for the probability of long-term storage of
the photon polarization q-bit is obtained and the conditions for the
photon storage with unit probability independent on its polarization
are outlined. The transitions with the angular momenta
$J_{b}=J_{c}=1$, $J_{a}=2$, corresponding to the transitions between
the hyperfine structure components of the electronic levels
$5^{2}S_{1/2}$ and $5^{2}P_{1/2}$ of the $^{87}Rb$ atom, which were
employed in the experiments \cite{n5}, are analyzed.

\section{Basic equations}

 \begin{figure}[t]\center
\includegraphics[width=7cm]{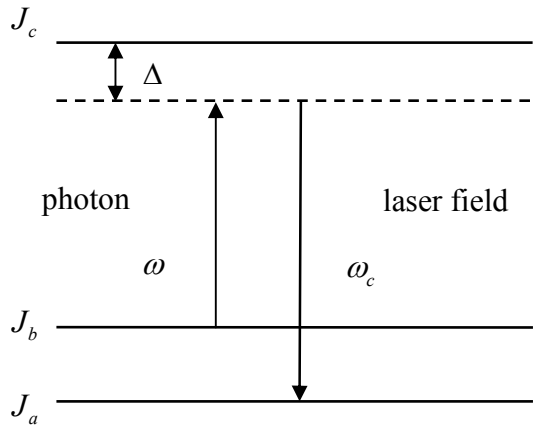}
\caption{The level diagram}
    \end{figure}

We consider the  three-level $\Lambda$ - type atom in a high-finesse
cavity, one branch -- the transition $J_{a}\rightarrow J_{c}$ with
the frequency $\omega_{c0}$ is coupled to the driving coherent laser
field, while the other -- the transition $J_{b}\rightarrow J_{c}$
with the frequency $\omega_{0}$ is coupled to the quantized cavity
field (Figure 1), $J_{a}$, $J_{b}$ and $J_{c}$ being the values of
the angular momenta of the levels. The electric field strength of
the coherent laser field may be put down as follows:
$$\textbf{E}_{c}=e_{c}(t)\textbf{l}_{c} e^{-i\omega_{c}t}+ c.c.,$$
while the quantized field of the cavity in the interaction
representation is described by the operator:
$$\hat{\textbf{E}}=e(t)(\hat{a}_{1}\textbf{l}_{1} +
\hat{a}_{2}\textbf{l}_{2}) e^{-i\omega t}+h.c.,$$ where the carrier
frequencies $\omega_{c}$ and $\omega$ are in resonance with the
frequencies $\omega_{c0}$ and $\omega_{0}$,  $e_{c}(t)$ and
$\textbf{l}_{c}$ are the slowly varying amplitude and the unit
polarization vector of the driving field, $e(t)$ is the slowly
varying amplitude of the photon field, $\textbf{l}_{1}$ and
$\textbf{l}_{2}$ are the two unit orthogonal vectors of the two
polarization modes of the photon field, $\hat{a}_{1}$ and
$\hat{a}_{2}$ are the photon annihilation operators for this modes.
The temporal dependence $e(t)$ of the cavity field amplitude appears
due to some tailored alteration of cavity parameters. The equation
for the slowly-varying density matrix $\hat{\rho}$ of the system,
which consists of a single three-level atom and two-mode cavity
field, in the rotating-wave approximation and in case of Raman
resonance $\omega_{c0}-\omega_{c}=\omega_{0}-\omega=\Delta$ is as
follows:
     \begin{equation}\label{q1}
 \frac{d}{dt} \hat{\rho} =
 \frac{i}{2}\left[\hat{V}(t),\hat{\rho}\right],
     \end{equation}
     \begin{equation}\label{q2}
\hat{V}(t) = -2\Delta\hat{P}_{c} + \hat{G}(t) + \hat{G}^{\dag}(t),
     \end{equation}
     \begin{equation}\label{q3}
\hat{G}(t) = \Omega_{a}(t)\hat{g}_{a} + \Omega_{b}(t)\hat{g}_{b},
     \end{equation}
     \begin{equation}\label{q4}
\hat{g}_{b}=\hat{g}_{b1}\hat{a}_{1}^{\dag}+
\hat{g}_{b2}\hat{a}_{2}^{\dag}.
    \end{equation}
Here $\hat{P}_{c}$ is the projector on the subspace of the atomic
excited level $J_{c}$, $\Omega_{a}(t)=2|d_{a}|e_{c}(t)/\hbar$ and
$\Omega_{b}(t)=2|d_{b}|e(t)/\hbar$ are the reduced Rabi frequencies
for the coherent laser field and for the cavity field,
$d_{a}=d(J_{a}J_{c})$ and $d_{b}=d(J_{b}J_{c})$ are the reduced
matrix elements of the electric dipole moment operator for the
transitions $J_{a}\rightarrow J_{c}$ and $J_{b}\rightarrow J_{c}$,
while
$$\hat{g}_{a}=\hat{\textbf{g}}_{a}\textbf{l}_{c}^{*},~
\hat{g}_{bi}=\hat{\textbf{g}}_{b}\textbf{l}^{*}_{i},~i=1,2,$$
$\hat{\textbf{g}}_{a}$ and $\hat{\textbf{g}}_{b}$ are the
dimensionless electric dipole moment operators for the transitions
$J_{a}\rightarrow J_{c}$ and $J_{b}\rightarrow J_{c}$. The matrix
elements of the circular components $\hat{g}_{\alpha q}$
($\alpha=a,b$; $q=0,\pm 1$) of these vector operators are expressed
through Wigner 3J-symbols \cite{n15}:
    \begin{equation}\label{q5}
  (\hat{g}_{\alpha q})_{m_{\alpha},m_{c}}=
(-1)^{J_{\alpha}-m_{\alpha}}\left(\matrix{J_{\alpha}&1&J_{c}  \cr
-m_{\alpha}&q&m_{c}}\right),
    \end{equation}
The solution of the equation (\ref{q1}) is expressed through the
evolution operator $\hat{S}(t)$:
    \begin{equation}\label{q6}
   \hat{\rho}(t)=\hat{S}(t)\hat{\rho}(0)\hat{S}^{+}(t).
    \end{equation}
In the adiabatic approximation the evolution operator $\hat{S}(t)$
is defined by the instantaneous eigenvectors $|v_{k}(t)>$ and
eigenvalues $\lambda_{k}(t)$ of the interaction operator
$\hat{V}(t)$ in a following way:
    \begin{equation}\label{q7}
\hat{S}(t)=\sum_{k}\exp\{i\phi_{k}(t)\} |v_{k}(t)><v_{k}(0)|,
    \end{equation}
    \begin{equation}\label{q8}
\phi_{k}(t)=\frac{1}{2}\int_{0}^{t}\lambda_{k}(t')dt'.
    \end{equation}
Since only single photon storage and retrieval are discussed in the
present paper and no decay or dephasing processes are taken into
account it is sufficient to limit the system space to the subspace
with the basis vectors $|J_{a}m_{a}>|0,0>$, $|J_{b}m_{b}>|1,0>$,
$|J_{b}m_{b}>|0,1>$ and $|J_{c}m_{c}>|0,0>$, where $|J_{a}m_{a}>$,
$|J_{b}m_{b}>$ and $|J_{c}m_{c}>$ denote the atomic Zeeman states,
while $|n_{1},n_{2}>$ ($n_{1,2}=0,1$) are the field number states
with $n_{1}$ photons in the first polarization mode and $n_{2}$ --
in the second. This subspace with the dimension
$N=2(J_{a}+2J_{b}+J_{c}+2)$ constitutes the invariant subspace of
the interaction operator $\hat{V}(t)$, so that its matrix represents
itself a square hermitian $N\times N$ matrix. The states $|a>|0,0>$
and $|c>|0,0>$, which atomic part belongs to the level $a$ or $c$,
may be represented as columns with $2J_{a}+1$ or $2J_{c}+1$ elements
correspondingly, while the states $|b_{1}>|1,0>+|b_{2}>|0,1>$, which
atomic part belongs to the level $b$, may be represented as columns
with $2(2J_{b}+1)$ elements:
$$|b_{1}>|1,0>+|b_{2}>|0,1> = \left(\matrix{|b_{1}> \cr
|b_{2}>}\right).$$ Then the operator $\hat{g}_{a}$ will be
represented by the $(2J_{a}+1)\times (2J_{c}+1)$ matrix, while the
operator $\hat{g}_{b}$ will be represented by the $2(2J_{b}+1)\times
(2J_{c}+1)$ matrix
$$\hat{g}_{b} = \left( \matrix{
\hat{g}_{b1} \cr \hat{g}_{b2}} \right),$$ were each block
$\hat{g}_{bi}$ ($i=1,2$) represents itself a $(2J_{b}+1)\times
(2J_{c}+1)$ matrix.

In order to find out the instantaneous eigenvectors $|v_{k}(t)>$ and
eigenvalues $\lambda_{k}(t)$ of the interaction operator
$\hat{V}(t)$ let us start with the operator
$$\hat{G}^{\dag}(t)\hat{G}(t) = \Omega_{a}^{2}(t)\hat{g}_{a}^{\dag}\hat{g}_{a}
+ \Omega_{b}^{2}(t)\hat{g}_{b}^{\dag}\hat{g}_{b},$$ which acts at
the subspace of the upper atomic level $c$ (note that
$\hat{g}_{a}^{\dag}\hat{g}_{b}=0$,
$\hat{g}_{b}^{\dag}\hat{g}_{a}=0$). The eigenvectors of this
operator form an orthonormal set, while its eigenvalues are real and
non-negative. The eigenvectors $|D_{n}^{c}(t)>$ with zero
eigenvalues:
    \begin{equation}\label{q9}
\hat{G}^{\dag}(t)\hat{G}(t)|D_{n}^{c}(t)>=0,~n=1,...,N_{c}^{d},
    \end{equation}
uncoupled to the lower levels $a$ and $b$, are at the same time the
eigenvectors of the interaction operator $\hat{V}(t)$ with the
eigenvalues $\lambda_{cn}=-2\Delta$. The eigenvectors $|C_{n}(t)>$
with positive eigenvalues $c_{n}^{2}(t)>0$:
    \begin{equation}\label{q10}
\hat{G}^{\dag}(t)\hat{G}(t)|C_{n}(t)>=c_{n}^{2}(t)|C_{n}(t)>,~
n=1,...,N_{c},
    \end{equation}
are coupled by the electric dipole transitions to the states
    \begin{equation}\label{q11}
|F_{n}(t)>=\frac{1}{c_{n}(t)}\hat{G}(t)|C_{n}(t)>.
    \end{equation}
These states $|F_{n}(t)>$ also form an orthonormal subset at the
subspace of the lower atomic levels $a$ and $b$, as it follows from
(\ref{q10})-(\ref{q11}).

With the introduction of the states $|C_{n}(t)>$ and $|F_{n}(t)>$
all the eigenvectors of the interaction operator $\hat{V}(t)$ with
non-zero eigenvalues may be easily obtained. This operator has
$2N_{c}$ eigenvectors
    \begin{equation}\label{q12}
|V_{n}^{(+)}(t)> = \sin \theta_{n}(t) |F_{n}(t) + \cos
\theta_{n}(t)|C_{n}(t)>,
    \end{equation}
    \begin{equation}\label{q13}
|V_{n}^{(-)}(t)> = \cos \theta_{n}(t) |F_{n}(t)> - \sin
\theta_{n}(t) |C_{n}(t)>,
    \end{equation}
    \begin{equation}\label{q14}
\tan 2\theta_{n}(t) = - \frac{c_{n}(t)}{\Delta},
    \end{equation}
with the eigenvalues
    \begin{equation}\label{q15}
\lambda_{n}^{(\pm)}(t) = -\Delta \pm \sqrt{\Delta^{2}+c_{n}^{2}(t)}.
    \end{equation}
All the states $|D_{n}^{c}(t)>$ and $|V_{n}^{(\pm)}(t)>$ constitute
the orthonormal set of $N^{f}=2(2J_{c}+1)-N^{d}_{c}$ eigenvectors of
the operator $\hat{V}(t)$ with non-zero eigenvalues.

The other $N^{d}=N-N^{f}$ eigenvectors $|D_{n}(t)>$ of the
interaction operator $\hat{V}(t)$  obtain zero eigenvalues. These
states -- dark states -- with the atomic part belonging to the
subspace of the lower atomic levels $a$ and $b$ remain uncoupled to
the upper level $c$:
    \begin{equation}\label{q16}
\hat{G}^{\dag}(t)|D_{n}(t)> = 0,~ n=1,...,N^{d}.
    \end{equation}
As it follows from (\ref{q11}) and (\ref{q16}), all the dark states
$|D_{n}(t)>$ are orthogonal to all the bright states $|F_{n}(t)>$.

Among all the dark states let us distinguish first the dark states
$|D_{n}^{a}>$ and $|D_{n}^{b}>$, which atomic part belongs to the
lower level $a$ or $b$ only. These states are time independent and
may be obtained as the eigenvectors of operators
$\hat{g}_{a}\hat{g}_{a}^{\dag}$ or $\hat{g}_{b}\hat{g}_{b}^{\dag}$
with zero eigenvalues:
     \begin{equation}\label{q17}
\hat{g}_{a}\hat{g}_{a}^{\dag}|D_{n}^{a}> = 0,~ k=1,...,N_{a}^{d},
     \end{equation}
     \begin{equation}\label{q18}
\hat{g}_{b}\hat{g}_{b}^{\dag}|D_{n}^{b}> = 0,~ k=1,...,N_{b}^{d}.
     \end{equation}

The other $N^{d}_{ab}$ eigenvectors $|D^{ab}_{n}(t)>$ of the
interaction operator $\hat{V}(t)$, which atomic part belongs to both
atomic lower levels $a$ and $b$, obtain zero eigenvalues and satisfy
the equation (\ref{q16}), while
$$\hat{g}_{a}^{\dag}|D_{n}^{ab}(t)> \neq 0,~
\hat{g}_{b}^{\dag}|D_{n}^{ab}(t)> \neq 0.$$  The temporal dependence
of these dark states may be immediately obtained from the equation
(\ref{q16}):
$$|D^{ab}_{n}(t)> = Z_{n}(t)[\Omega_{a}(t)|\tilde{B}_{n}^{d}> -
\Omega_{b}(t)|A_{n}^{d}>],$$ where $Z_{n}(t)$ is the normalization
factor, while $|A_{n}^{d}>$ and $|\tilde{B}_{n}^{d}>$ are temporally
independent states, which atomic parts belong to the levels $a$ and
$b$ correspondingly and which satisfy the equation
    \begin{equation}\label{q19}
\hat{g}_{a}^{\dag}|A_{n}^{d}> =
\hat{g}_{b}^{\dag}|\tilde{B}_{n}^{d}>\neq 0.
    \end{equation}
Introducing the matrix
$$\hat{D}_{b} = \sum_{n}\frac{1}{b_{n}^{2}}
|B_{n}><B_{n}|,$$ containing only the eigenvectors $|B_{n}>$ of
matrix $\hat{g}_{b}\hat{g}_{b}^{\dag}$ with non-zero eigenvalues
$b_{n}^{2}$, we may write the equation (\ref{q19}) as follows:
    \begin{equation}\label{q20}
|\tilde{B}_{n}^{d}> = \hat{D}_{ba}|A_{n}^{d}>,~ \hat{D}_{ba} =
\hat{D}_{b}\hat{g}_{b}\hat{g}_{a}^{\dag}.
    \end{equation}
Now we may define the orthonormal set of states $|A_{n}^{d}>$ as the
eigenvectors of the hermitian matrix
$\hat{D}_{ba}^{\dag}\hat{D}_{ba}$ with non-zero eigenvalues:
    \begin{equation}\label{q21}
\hat{D}_{ba}^{\dag}\hat{D}_{ba}|A_{n}^{d}> = a_{dn}^{2}|A_{n}^{d}>,~
a_{dn}>0.
    \end{equation}
Then, as it follows from (\ref{q20}) and (\ref{q21}), the states
   \begin{equation}\label{q22}
|B_{n}^{d}> = \frac{1}{a_{dn}}\hat{D}_{ba}|A_{n}^{d}>
    \end{equation}
also constitute the orthonormal set, so that the orthonormal set of
dark states $|D^{ab}_{n}(t)>$ may be expressed through these states
$|A_{n}^{d}>$ and $|B_{n}^{d}>$ by the formula:
    \begin{equation}\label{q23}
|D^{ab}_{n}(t)> = \frac{a_{dn}\Omega_{a}(t)|B_{n}^{d}> -
\Omega_{b}(t)|A_{n}^{d}>}
{\sqrt{\Omega_{b}^{2}(t)+a_{dn}^{2}\Omega_{a}^{2}(t)}}.
    \end{equation}
All the eigenvectors of the interaction operator $\hat{V}(t)$,
comprising the set of states $|D_{n}^{c}(t)>$,
 $|V_{n}^{(\pm)}(t)>$, and the set of dark
states $|D_{n}^{a}>$, $|D_{n}^{b}>$ and $|D_{n}^{ab}(t)>$ constitute
the complete orthonormal set of states, which determines the
evolution operator (\ref{q7}) in the adiabatic approximation.

\section{Photon storage and retrieval}

Initially the atom is at the lower level $b$, its state being
defined by the atomic density matrix $\hat{\rho}_{0}^{b}$, while the
state of the field is a pure single-photon state
$$|f>=\xi_{1}|1,0>+\xi_{2}|0,1>,~|\xi_{1}|^{2}+|\xi_{2}|^{2}=1,$$
where the two complex numbers $\xi_{1}$ and $\xi_{2}$ define the
photon polarization (the polarization q-bit), so that the initial
density matrix $\hat{\rho}(0)$ of the atom+field system is:
$$\hat{\rho}(0) = \hat{\rho}_{0}^{b}\cdot|f><f|.$$
The classical coherent laser field is adiabatically switched off,
while the interaction with the quantum field is adiabatically
switched on in the time interval $T_{1}$, so that:
$$\Omega_{a}(0)=\Omega_{a1},~ \Omega_{a}(T_{1})=0,
\Omega_{b}(0)=0,~ \Omega_{b}(T_{1})=\Omega_{b1}.$$ The atomic part
of the initial bright states $|F_{n}(0)>$, defined by the equation
(\ref{q11}), with non-zero eigenvalues $c^{2}_{n}(0)\neq 0$ belongs
to the level $a$ only, so that the corresponding initial
eigenvectors $|V^{\pm}_{n}(0)>$ of the interaction operator, defined
by (\ref{q12})-(\ref{q14}), will not contribute to the evolution.
However the atomic part of the initial bright states $|F_{n}(0)>$
with zero eigenvalues $c^{2}_{n}(0)=0$, obtained as the limit at
$t\rightarrow +0$, belongs to the level $b$. Since
$c^{2}_{n}(0)=\Omega_{a1}^{2}c^{2}_{an}$, where $c^{2}_{an}$ are the
eigenvalues of the operator $\hat{g}_{a}^{\dag}\hat{g}_{a}$, such
states will be coupled to the eigenvectors $|C_{n}^{b}>$ of the
operator $\hat{g}_{a}^{\dag}\hat{g}_{a}$ with zero eigenvalues:
   \begin{equation}\label{q24}
\hat{g}_{a}^{\dag}\hat{g}_{a}|C_{n}^{b}>=0,~n=1,...,N_{c}^{b}.
    \end{equation}
So, the evolution operator (\ref{q7}) will be determined by the
normalized states $\hat{g}_{b}|C_{n}^{b}>$ and the dark states
$|D_{n}^{b}>$ and $|D_{n}^{ab}(t)>$, defined by (\ref{q18}) and
(\ref{q23}), which atomic part also belongs to the level $b$.  The
states $|D_{n}^{b}>$ in the evolution operator are responsible for
the photon passage through the cavity without interaction, while the
states $\hat{g}_{b}|C_{n}^{b}>$ are responsible for the process of
photon absorption and the adiabatic transfer of the atom from the
initial level $b$ to the excited level $c$ to some state, which is
uncoupled to the lower level $a$. In both cases the final system
state is unstable and for purposes of long-term photon storage the
presence of such states in the initial system state should be
avoided by the proper choice of polarization of the driving field
and by the preparation of the atomic state. With such states
excluded the evolution is determined only by the states
$|D_{n}^{ab}(0)>$, which are responsible for the long-term recording
of the photon polarization q-bit onto the superposition of Zeeman
substates of the metastable level $a$. The evolution operator at the
instant of time $t=T_{1}$ after the recording process is finished is
as follows:
   \begin{equation}\label{q25}
\hat{S}_{1}=-\sum_{n}|A_{n}^{d}><B_{n}^{d}|,
    \end{equation}
so that each state $|B_{n}^{d}>$ from the level $b$, defined by
(\ref{q22}), is adiabatically transferred to the corresponding state
$|A_{n}^{d}>$ from the level $a$, defined by (\ref{q21}). The total
probability $w$ of the photon storage at the level $a$ generally
depends on the photon polarization $\xi_{1}$ and $\xi_{2}$:
    \begin{equation}\label{q26}
w= tr\{\hat{S}_{1}\hat{\rho}(0)\hat{S}_{1}^{\dag}\} =
\sum_{i,j=1}^{2}\xi_{i}\xi_{j}^{*}w_{ij},
    \end{equation}
where
    \begin{equation}\label{q27}
w_{ij}=\sum_{n}<b_{n}^{di}|\hat{\rho}_{0}^{b}|b_{n}^{dj}>,
    \end{equation}
while $|b_{n}^{d1}>=<1,0|B_{n}^{d}>$ and
$|b_{n}^{d2}>=<0,1|B_{n}^{d}>$ are the pure atomic states at the
level $b$ -- the projections of the field+atom states $|B_{n}^{d}>$
on the two photon polarization states $|1,0>$ and $|0,1>$. For
purposes of q-bit storage the total probability must be unity and
must not depend on photon polarization, so that the condition
    \begin{equation}\label{q28}
w_{ij}=\delta_{ij}
    \end{equation}
must be satisfied.

To retrieve the photon after the storage time $\tau$ the classical
coherent laser field with the same polarization $\textbf{l}_{c}$, as
that of the field applied to store the photon, is adiabatically
switched on, while the interaction with the quantum field is
adiabatically switched off in the time interval $T_{2}$, so that:
 $$\Omega_{a}(T_{1}+\tau)=0,~
 \Omega_{a}(T_{1}+\tau+T_{2})=\Omega_{a2},$$
$$\Omega_{b}(T_{1}+\tau)=\Omega_{b2},~
\Omega_{b}(T_{1}+\tau+T_{2})=0.$$ If the photon was stored with unit
probability, then at the start of storage process the atom was
prepared at some state of the level $b$, which contained only the
parts $|B_{n}^{d}>$ of the "dark" states $|D_{n}^{ab}>$, while the
the initially unpopulated level $a$ was populated during the storage
process by the adiabatic transfer of the parts $|B_{n}^{d}>$ of the
"dark" states $|D_{n}^{ab}>$ to their counterparts $|A_{n}^{d}>$ at
the level $a$, so that at the start of the retrieval stage only the
"dark" states $|D_{n}^{ab}>$ will be present in the initial atomic
state, provided that the polarization of the retrieving field is the
same as that of the recording field, ensuring that the "dark" states
$|D_{n}^{ab}>$ are the same for both stages -- the storage and the
retrieval. In this case only the dark states $|D_{n}^{ab}(t)>$ will
contribute to the evolution operator (\ref{q7}). Then, at the end of
the retrieval process at the instant of time
$t_{2}=T_{1}+\tau+T_{2}$ the evolution operator
    \begin{equation}\label{q29}
\hat{S}_{2}=-\sum_{n}|B_{n}^{d}><A_{n}^{d}|
    \end{equation}
is determined by the same states $|A_{n}^{d}>$ and $|B_{n}^{d}>$ as
the evolution operator $\hat{S}_{1}$ (\ref{q25}) and the product of
the evolution operators $\hat{S}_{2}\hat{S}_{1}$ is reduced to the
projector on the subspace of states $|B_{n}^{d}>$:
    $$\hat{S}_{2}\hat{S}_{1}=\sum_{n}|B_{n}^{d}><B_{n}^{d}|,$$
and since only such states are present at the initial system state
$\hat{\rho}(0)$, then this state will be retrieved without
distortion
    \begin{equation}\label{q30}
\hat{\rho}_{2}=\hat{S}_{2}\hat{S}_{1} \hat{\rho}(0)
\hat{S}^{\dag}_{2}\hat{S}^{\dag}_{1} = \hat{\rho}(0),
    \end{equation}
reproducing a single photon $|f>$ with the recorded polarization and
the atom at the initial state with the density matrix
$\hat{\rho}_{0}^{b}$.

Let us choose the two unit polarization vectors $\textbf{l}_{1}$ and
$\textbf{l}_{2}$ of the quantum field in the plane $XY$ as the two
vectors with opposite circular components:
$$l_{1q} = \delta_{q,-1},~ l_{2q} = \delta_{q,1},$$
while the polarization vector $\textbf{l}_{c}$ of the classical
field generally contains also $Z$-projections.

In the experiment \cite{n5} the angular momenta of the atomic levels
were $J_{a}=2$, $J_{b}=J_{c}=1$, while the driving field was
linearly polarized along the axis $Z$: $l_{cq} =\delta_{q,0}$
($\pi$-polarized field). In this case there are three states
$|A_{n}^{d}>$ at the level $a$, defined by (\ref{q21}):
$$ |A_{1}^{d}> = \left( \matrix{0\cr 0\cr 0\cr 1\cr 0}\right),~
|A_{2}^{d}> = \left( \matrix{0\cr 1\cr 0\cr 0\cr 0}\right),~
|A_{3}^{d}> = \left( \matrix{0\cr 0\cr 1\cr 0\cr 0}\right),$$ which
are adiabatically coupled to the three states $|B_{n}^{d}>$ at the
level $b$, defined by (\ref{q22}):
$$ |B_{1}^{d}> = \left( \matrix{0 \cr 1 \cr 0}\right)|1,0>,~
|B_{2}^{d}> = -\left( \matrix{0 \cr 1 \cr 0}\right)|0,1>,$$
$$|B_{3}^{d}> = \frac{1}{\sqrt{2}}\left\{\left( \matrix{1 \cr 0 \cr
0}\right)|1,0>-\left( \matrix{0 \cr 0 \cr 1}\right)|0,1>\right\}.$$
The elements of matrix $w_{i,j}$ (\ref{q27}) determining the photon
storage probability are expressed through the elements
$\rho^{bb}_{m_{b},m'_{b}}$ of the initial atomic density matrix
$\hat{\rho}_{0}^{b}$ in a following way:
$$w_{1,1} = \rho^{bb}_{0,0} + \frac{1}{2}
\rho^{bb}_{-1,-1},$$ $$w_{2,2} = \rho^{bb}_{0,0} + \frac{1}{2}
\rho^{bb}_{1,1},$$ $$w_{2,1} = w_{1,2}^{*} = - \frac{1}{2}
\rho^{bb}_{1,-1},$$ so that the conditions (\ref{q28}) may be met
only if the atom is initially in the pure state $|J_{b}=1,m_{b}=0>$,
as it was in the experiment. Then, as it follows from the expression
(\ref{q25}) for the evolution operator, the initial system pure
state
$$|J_{b}=1,m_{b}=0>(\xi_{1}|1,0>+\xi_{2}|0,1>)$$
is adiabatically transferred to the pure state
$$(\xi_{2}|J_{a}=2,m_{a}=-1> - \xi_{1}|J_{a}=2,m_{a}=1>)|0,0>.$$
The conditions (\ref{q28}) may be also met, as calculations show,
with the driving field linearly polarized along the axis $X$:
$$l_{cq}=\frac{1}{\sqrt{2}}\left(\delta_{q,-1}-\delta_{q,1}\right)$$
and with the same initial atomic state $|J_{b}=1,m_{b}=0>$, though
in this case the photon polarization will be mapped to the
superposition of states $|J_{a}=2,m_{a}=0>$ and $|J_{a}=2,m_{a}=\pm
2>$:
$$a^{0}|m_{a}=0> + a^{+}|m_{a}=2> + a^{-}|m_{a}=-2>,$$
$$a^{0}=\frac{\xi_{1}+\xi_{2}}{2\sqrt{2}},~
a^{\pm}=\frac{\sqrt{3}(\xi_{1}+\xi_{2})}{4} \pm
\frac{(\xi_{1}-\xi_{2})}{2}.$$ For the lower values $J_{b}=0$ and
$J_{b}=1/2$  of the angular momenta of the initial atomic level $b$
the conditions (\ref{q28}) for the photon storage with unit
probability, independent on photon polarization, may be easily met.
With the driving field linearly polarized along the axis $Z$:
$l_{cq} =\delta_{q,0}$, these conditions will be satisfied, for
example, for the transitions with $J_{b}=0$, $J_{a}=J_{c}=1$ and
with $J_{b}=1/2$, $J_{a}=J_{c}=3/2$ for any initial atomic state, in
particular, for the equilibrium state: $\rho^{bb}_{-1/2,-1/2} =
\rho^{bb}_{1/2,1/2}=1/2$, $\rho^{bb}_{-1/2,1/2} =
\rho^{bb}_{1/2,-1/2}=0$. However for larger values of angular
momentum $J_{b}>1$ it becomes hard (if not impossible) to meet the
conditions (\ref{q28}). Indeed, the memory efficiency (or better to
say inefficiency) is determined by the number $N_{d}^{b}$ of the
"dark" states of the type $|D_{n}^{b}>$, defined by the equation
(\ref{q18}), and by the weight of these states in the initial atomic
state. The number of states $2(2J_{b}+1)$ with the atomic part at
the level $b$ is doubled due to the two possible polarizations of
the cavity field and only $2J_{c}+1$ of them are coupled to the
excited level $c$, while the other $N_{d}^{b}=2(2J_{b}-J_{c}+1)-1$
states -- the "dark" states of the type $|D_{n}^{b}>$ -- stay
uncoupled, reducing the memory efficiency. The number of such states
obtains its minimum $N_{d}^{b}=2J_{b}-1$ on the transitions
$J_{b}\rightarrow J_{c}=J_{b}+1$, so that at $J_{b}\geq 1$ the
presence of these states becomes mandatory. The number $N_{d}^{b}$
of the states $|D_{n}^{b}>$ grows with $J_{b}$ and at $J_{b}>1$ it
becomes highly improbable that each Zeeman state $|J_{b},m_{b}>$
will not be contained in these "dark" states $|D_{n}^{b}>$. The
general approach, developed in the present paper, enables one to
analyze numerically the memory efficiency for any reasonable values
of the angular momenta of resonant atomic levels. For example, let
us consider the transitions $J_{b}=2\rightarrow J_{c}=3\rightarrow
J_{a}=4$ and the $\pi$-polarized driving field ($l_{cq}
=\delta_{q,0}$). Then the elements of the probability matrix
$w_{i,j}$ (\ref{q27}) are expressed through the elements
$\rho^{bb}_{m_{b},m'_{b}}$ of the initial atomic density matrix
$\hat{\rho}_{0}^{b}$ in a following way:
      \begin{equation}\label{q31}
w_{1,1} = \frac{1}{7}\rho^{bb}_{-2,-2} +
\frac{1}{2}\rho^{bb}_{-1,-1} + \frac{6}{7}\rho^{bb}_{0,0} +
\rho^{bb}_{1,1} + \rho^{bb}_{2,2},
       \end{equation}
       \begin{equation}\label{q32}
w_{2,2} = \frac{1}{7}\rho^{bb}_{2,2} + \frac{1}{2}\rho^{bb}_{1,1} +
\frac{6}{7}\rho^{bb}_{0,0} + \rho^{bb}_{-1,-1} + \rho^{bb}_{-2,-2},
       \end{equation}
while $w_{1,2}$ and $w_{2,1}$ are expressed only through the
non-diagonal elements $\rho^{bb}_{m_{b},m'_{b}}$ ($m_{b}\neq
m'_{b}$) of the atomic density matrix. In addition  the populations
of Zeeman sublevels are non-negative $\rho^{bb}_{mm}\geq 0$ and the
total population of the level is unity
      \begin{equation}\label{q33}
\rho^{bb}_{-2,-2} + \rho^{bb}_{-1,-1} + \rho^{bb}_{0,0} +
\rho^{bb}_{1,1} + \rho^{bb}_{2,2} = 1.
       \end{equation}
For a hundred per cent memory efficiency $w_{1,1}=w_{2,2}=1$, then
extracting (\ref{q31}) from (\ref{q33}) we obtain
$$\frac{6}{7}\rho^{bb}_{-2,-2} + \frac{1}{2}\rho^{bb}_{-1,-1} +
\frac{1}{7}\rho^{bb}_{0,0} = 0,$$ which is true only if
$\rho^{bb}_{-2,-2} = \rho^{bb}_{-1,-1} = \rho^{bb}_{0,0} = 0$, and
extracting (\ref{q32}) from (\ref{q33}) we also obtain that
$\rho^{bb}_{2,2} = \rho^{bb}_{1,1} = 0$, so that in this case the
photon q-bit cannot be stored with unit probability. However, if the
atom is initially prepared at the state $|J_{b}=2,m_{b}=0>$, then
$w_{ij}=(6/7)\delta_{ij}$, and the photon will be stored with the
probability $w=6/7$ independent of its polarization.

\section{Conclusions}

In the present paper we have obtained the general expression for the
probability of long-term storage of a single-photon polarization
q-bit at the magnetic substates of some long-lived degenerate atomic
level by means of STIRAP. The dependence of this probability on the
polarization of the driving field and on the initial atomic state
was studied. It was shown that the conditions for the photon
long-term storage with unit probability independent on the photon
polarization may be satisfied with low values of angular momentum of
the initial atomic level $J_{b}=0,1/2,1$. At larger values of this
angular momentum $J_{b}>1$ it becomes hard (if not impossible) to
store the arbitrarily polarized photon with unit probability. The
general approach, developed in the present paper, enables one to
analyze numerically the memory efficiency for any reasonable values
of the angular momenta of resonant atomic levels. It was also shown
that, since the conditions of the photon storage with unit
probability are satisfied, the driving field with the same
polarization as that used to store the photon will retrieve the
initial atomic state and a single photon with the recorded
polarization without distortion.

{\bf Acknowledgements}

Authors are indebted for financial support of this work to Russian
Ministry of Science and Education (grant 2407).

\end{document}